\documentclass{ws-procs975x65}

\usepackage{graphicx}

\begin{document}

\title{CHAOS IN CORE-HALO GRAVITATING SYSTEMS}

\author{T. GHAHRAMANYAN$^1$, V.G. GURZADYAN$^{1,2,*}$}

\address{$^1$Yerevan Physics Institute, Yerevan, Armenia; $^2$ICRANet, Dipartimento
di Fisica, Universita La Sapienza, Roma, Italy\\
$^*$E-mail: gurzadya@icra.it}

\begin{abstract}
Chaotic dynamics essentially defines the global properties of gravitating systems, including, probably, the basics of morphology of galaxies. We use the Ricci curvature criterion to study the degree of relative chaos (exponential instability) in core-halo gravitating configurations. We show the existence of a critical core radius when the system is least chaotic, while systems with both smaller and larger core radius will typically possess stronger chaotic properties.
\end{abstract}

\keywords{Chaos; gravitating systems; galaxies.}

\bodymatter

\section{Introduction}

The importance of chaotic effects in the dynamics of N-body gravitating systems has been attracting attention during the last decades\cite{GS,P,GP,GIAU,B} . The difficulty of rigorous study of chaos in 3D gravitating N-body systems is partly determined by the limited content of such descriptors as Lyapunov characteristic exponents, otherwise applicable for low degree of freedom systems. Numerous numerical studies estimating not clearly defined Lyapunov-like exponents remain not helpful in deciphering the complex nature and far going consequences of chaos in many dimensional nonlinear systems (see the critics in\cite{Ru}).

Below we represent a brief summary of the study of statistical properties of core-halo type gravitating systems\cite{LB}, using the Ricci curvature criterion of relative instability. That criterion was introduced in\cite{GK} upon discussions with Vladimir Arnold, and later was applied in extensive numerical studies of N-body systems, see e.g.\cite{EZ}. 
We have investigated spherical stellar systems with a core of various radii, following the behavior of the Ricci curvature depending on a ratio of core $r_c$ and system's $R$ radii, $k = r_c / R$. That dependence is also observed while varying the total energy of the system. 

\section{The Ricci Criterion}

Well known geometric methods of theory of dynamical systems enable one to study the properties of a Hamiltonian system reducing the equations of motion to a geodesic flow in the configuration space $M$ \cite{Arn,GP}. For further development of these methods see\cite{K}. 
The Ricci curvature $r_u(s)$ in $M$ in the direction of the velocity vector $u$ of the geodesics is defined as
\begin{equation}
r_u(s) = R_{ij}u^i u^j/||u||^2.
\end{equation}
Averaged with respect to the set of perturbed N-body systems,  
within an interval $[0, s_*]$ of the affine parameter of the geodesics, it will yield 
$$
r=\frac{1}{3N}\displaystyle\inf_{0 \leq s \leq s_*} {r_u(s)}.
$$ 
At smaller negative values of $r_u$ a system is unstable with higher probability, as the deviation vector $z(s)$ of the close geodesics increases faster
\begin{equation}
z(s) \geq e^{\sqrt {-r} s},
\end{equation}
in that interval.
For collisionless N-body systems $r_u(s)$ is\cite{GK}
\begin{equation}
r_u(s) = 
-\frac{3N - 2}{2} \frac {W_{i,k} u^i u^k} {W} 
+ \frac{3}{4}(3N - 2) \frac {(W_i u^i)^2} {W^2} 
- \frac{3N-4}{4} \frac {|\bigtriangledown W|^2} {W^3},
\end{equation}
where $W = E - V(q)$, $V(q)$ is the Newtonian potential, and $E$ is the total energy of the system, $W_i$ are the derivatives of $W$.

\section{Numerical Analysis}

Spherical 3D systems have been created with randomly generated velocities and coordinates of N point particles of equal mass. Each system consisted of two concentric spheres, initially both spheres having the same radius $k=1$, then with appearance of a core by means of the decrease of $k$. To enable the comparison of the sequence of the created systems, the total energy parameter has been fixed via the multiplication of all velocities of the system by certain constants. The estimation of the Ricci for such static configurations describes the role of the core in the instability properties of the system immediately moving away from the initial time moment. Typical behavior of the Ricci curvature is exhibited in Figure 1 vs $k$, and the total energy as a parameter. 

\begin{figure}[thb]%
\begin{center}
\psfig{file=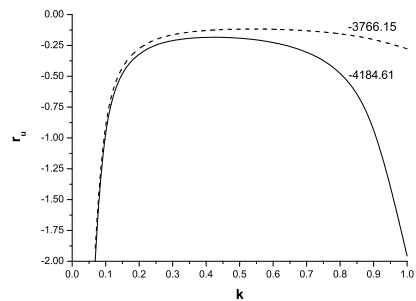,width=2.4in}
\caption{The dependence of the Ricci curvature on the ratio of the core and system's radii, $k$, for two values of total energy of the system for N=1000.}
\end{center}
\end{figure}

One can see that, the Ricci curvature has a maximum at some value of $k=k_{cr}$. The latter corresponds to the most stable system among those with different core radii, so that for both $k \rightarrow 0$ and $k \rightarrow 1$, the system becomes more unstable; we know that spherical N-body systems are exponentially instable as Kolmogorov systems\cite{GS}. The value of $k_{cr}$ has been investigated for different systems, varying the total energy, the number of stars and the radius of the system. The behavior of $k_{cr}$ vs the total (negative) energy $E$ is shown in Figure 2.

\begin{figure}[thb]%
\begin{center}
\psfig{file=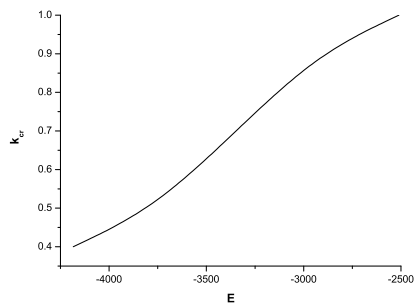,width=2.4in}
\caption{The variation of $k_{cr}$ vs the total energy of the system $E$ for N=1000.}
\end{center}
\end{figure}

Core-halo configurations are typical for the observed stellar systems, globular clusters and elliptical galaxies, and were an object of numerous theoretical studies, including the pioneering one by Lynden-Bell \cite{LB}. Note, we do not discuss core collapse type evolutionary effects, as they have much larger characteristic time scales than those of the reaching the quasi-stationary states discussed here. 
The existence of a critical core radius as revealed above, can bring closer the link with thermodynamic and other approaches to the stability of spherical gravitating systems.

\bibliographystyle{ws-procs975x65}
\end{document}